\documentclass[aps,prd,twocolumn,superscriptaddress,showpacs,letter,nofootinbib]{revtex4-1}
\usepackage{graphicx}
\usepackage{dcolumn}
\usepackage{bm}
\usepackage{cancel}
\usepackage{color}

\usepackage{amssymb}
\newcommand{\ba}{\begin{array}}
\newcommand{\ea}{\end{array}}

\def\br{\begin{eqnarray}}
\def\er{\end{eqnarray}}
\def\be{\begin{equation}}
\def\ee{\end{equation}}

\def\({\left(}
\def\){\right)}

\begin{document}
      
\title{Semihard interactions at TeV energies}

\author{T.~V.~Iser}
\email{iser@ufrgs.br}
\affiliation{Instituto de F\'isica, Universidade Federal do Rio Grande do Sul, Caixa Postal 15051, 91501-970, Porto Alegre, RS, Brazil}
\author{E.~G.~S.~Luna}
\email{luna@if.ufrgs.br}
\affiliation{Instituto de F\'isica, Universidade Federal do Rio Grande do Sul, Caixa Postal 15051, 91501-970, Porto Alegre, RS, Brazil}


\begin{abstract}

We investigate the high-energy behavior of the total cross section, $\sigma_{\text{tot}}$, and the ratio of the real to imaginary parts of the scattering amplitude, $\rho$, in both proton-proton and antiproton-proton channels. Our analysis is based on a QCD-inspired model in which the rise of the cross sections is predominantly driven by semihard processes involving gluons.
We address the tension between measurements from the ATLAS/ALFA and TOTEM Collaborations, showing that independent analyses of their datasets can provide statistically consistent descriptions of the overall data, even though they do not fully reproduce the central values of the $\rho$ parameter at $\sqrt{s} = 13$ TeV. 
The slight discrepancy between these central values and the model's predicted values, obtained using an asymptotic dominant crossing-even elastic scattering amplitude, points to the potential presence of an odd component in the semihard amplitude at high energies.

\end{abstract}


\maketitle

\section{Introduction}

According to QCD, the rise of the total cross section with energy in hadronic collisions is driven by jets with transverse energy $E_{T}$ that is much smaller than the square of the total center-of-mass energy $s$ involved in the collision \cite{evidences001,evidences001a,evidences001b,evidences001c,evidences001d}. These {\it minijets} originate from semihard scatterings of partons, which are hard scatterings of elementary partons carrying very small fractions of momenta of their parent hadrons \cite{ryskin001,ryskin001a,ryskin001b,ryskin002}. Their production is expected to dominate hadronic interactions at very high energies as jet-containing events become increasingly abundant with rising energy. Since jets and minijets result from hard and semihard partonic processes, minijet models assume that semihard dynamics play a central role in hadronic collisions at high energies.

From a phenomenological standpoint, the rise of total cross sections can be described within an eikonal QCD-based framework that respects both analyticity and unitarity constraints \cite{qcdmodel001,qcdmodel001a,qcdmodel001b,qcdmodel002,qcdmodel002a,qcdmodel002b,luna001,luna001a,iser01}. Specifically, the energy dependence of $\sigma_{\text{tot}}(s)$ and $\rho(s)$ can be derived from the QCD parton model by using standard parton-parton cross sections, updated sets of parton distribution functions (PDFs) and cutoffs which restrict the parton-level interactions to the semihard regime. In this picture, hadronic scattering emerges as an incoherent summation over all possible scatterings of constituent partons.

At ultrahigh energies, extrapolations typically assume that the eikonal function, which models the hadronic interaction, can be decomposed into two components: one associated with semihard (minijet) interactions and another representing purely soft processes. The soft eikonal term requires a separate, independent model in the soft limit, where the hadron behaves as a coherent system during scattering.

This work revisits the minijet-model formalism in order to describe the total cross section and the $\rho$ parameter measured at LHC energies. Specifically, we employ a minijet-based approach to compute these forward observables at high energies, where hard and semihard processes are expected to contribute significantly to the elastic scattering amplitude. Our analysis also addresses the tension between the measurements reported by the ATLAS/ALFA and TOTEM Collaborations. By introducing normalization factors associated with uncertainties in the luminosity determination, we show that it is possible to achieve a coherent global description of the data from each collaboration.

Finally, from a statistical perspective, we show that the $\rho$ data at 13 TeV can be reasonably described within a model where the scattering amplitude is dominated solely by crossing-even elastic terms, although the model cannot fully account for their central values.\footnote{In Regge theory, the asymptotic crossing-even contribution corresponds to the Pomeron, a colorless state having the quantum numbers of the vacuum.} This result suggests that an accurate description of the central values of the $\rho$ parameter at 13 TeV may require a scattering amplitude that includes an asymptotically surviving odd-under-crossing term.\footnote{The Odderon is the $C=-1$ partner of the $C=+1$ Pomeron. In the QCD language, the Odderon can be associated to a colorless $C_{odd}$ $t$-channel state with an intercept at or near one \cite{ewerz001,ewerz001a,ewerz001b,ryskin01}.}

The outline of this paper is as follows. Sec. II provides a review of the eikonal minijet formalism, along with the unitarity and analyticity properties of the scattering amplitude. In Sec. III we introduce our model, where an even-under-crossing term primarily governs the asymptotic scattering amplitude.
In Sec. IV we present our results and conclusions, exploring the tension between the TOTEM and ATLAS/ALFA measurements, and examining the impact of different PDFs on the behavior of $\sigma_{\text{\text{tot}}}$ and $\rho$ at high energies.

\section{Formalism}

A physically well-grounded calculation of high-energy hadron-hadron cross sections must respect the constraints imposed by analyticity and unitarity. In the impact parameter, $b$, representation, the unitarity condition implies the $s$-channel two-particle unitarity equation \cite{barone001,forshaw001}
\begin{eqnarray}
2 \textnormal{Re}\, H(s,b) = |H(s,b)|^{2} + G_{inel}(s,b) ,
\label{unitarity001}
\end{eqnarray}
where $G_{\text{inel}}(s,b)$ is a real, non-negative quantity accounting for contributions from inelastic channels, and $H(s,b)$ is the elastic scattering amplitude in $b$-space. The amplitude $H(s,b)$ is related to the momentum-space elastic scattering amplitude, $\mathcal{A}(s,t)$, through the relation
\begin{eqnarray}
{\cal A}(s,t) = i \int_{0}^{\infty}b\, db\, J_{0}(b\sqrt{-t})\, H(s,b),
\label{scatter001}
\end{eqnarray}
where $q = \sqrt{-t}$ is the momentum transfer and $J_0(x)$ is the Bessel function of the first kind. The unitarity condition can be naturally satisfied by using an eikonal approach to describe hadronic interactions, where we write $H(s,b) = 1 - e^{-\chi(s,b)}$. Here, the eikonal function $\chi(s,b)$ is a complex function: $\chi(s,b)=\textnormal{Re}\, \chi(s,b) + i\textnormal{Im}\, \chi(s,b)\equiv\chi_{_{R}}(s,b)+i\chi_{_{I}}(s,b)$. For the scattering process between two hadrons $A$ and $B$, the elastic amplitude ${\cal A}_{_{AB}}(s,t)$, expressed in terms of $\chi_{_{AB}}(s, b)$, becomes
\begin{eqnarray}
{\cal A}_{_{AB}}(s,t) = i \int_{0}^{\infty} b\, db\, J_{0}(b\sqrt{-t}) \left[1-e^{-\chi_{_{AB}}(s,b)} \right];
\label{equation02}
\end{eqnarray}
thus, once the eikonal function is known, the scattering amplitude is fully determined.

The analyticity of the scattering amplitude ${\cal A}(s,t)$ gives rise to dispersion relations that incorporate the condition of crossing symmetry. For elastic processes in the forward direction ($t=0$), the crossing variable corresponds to the energy $E$ of the incident particle in the laboratory frame. Denoting by ${\cal F}(E)$ the analytic continuation of the forward elastic scattering amplitude, ${\cal A}(E,t=0)$, the forward amplitudes for $pp$ (proton-proton) and $\bar{p}p$ (antiproton-proton) scattering are then obtained in the respective limits
\begin{eqnarray}
{\cal A}_{pp}(E,t=0) = \lim_{\epsilon \to 0} {\cal F}( E + i\epsilon, t=0) ,
\end{eqnarray}
\begin{eqnarray}
{\cal A}_{\bar{p}p}(E,t=0) = \lim_{\epsilon \to 0} {\cal F}(- E - i\epsilon, t=0) ,
\end{eqnarray}
where $E=(s-2m^{2})/2m$. As a result, the functions
\begin{eqnarray}
{\cal A}^{\pm}(E,t=0) = \frac{1}{2} \left[ {\cal A}_{\bar{p}p}(E,t=0) \pm  {\cal A}_{pp}(E,t=0)  \right]
\end{eqnarray}
are found to be even ($+$) and odd ($-$) real analytic functions of $E$ that satisfy dispersion relations.

In the limit $E \gg m$, by changing the variable from $E$ to $s$ and decomposing the eikonal function into its even and odd components, namely $\chi^{\bar{p}p}_{pp}(s,b)=\chi^{+}(s,b)\pm\chi^{-}(s,b)$, we obtain the relations
\begin{eqnarray}
{\cal A}^{+}(s,0) = i \int_{0}^{\infty}db \, b\, \left[ 1 - \cosh\left( \chi^{-} \right) \exp \left(-\chi^{+} \right)\right] ,
\label{cut1}
\end{eqnarray}
\begin{eqnarray}
{\cal A}^{-}(s,0) = i \int_{0}^{\infty}db \, b\, \sinh\left( \chi^{-} \right) \exp \left(-\chi^{+} \right) .
\label{cut2}
\end{eqnarray}

We can therefore operate directly with the eikonals rather than the amplitudes, as illustrated by relations (\ref{cut1}) and (\ref{cut2}). These relations indicate that by assuming $\chi^{+}$ and $\chi^{-}$ are, respectively, even and odd real analytic functions of $E$ with the same cut structure as ${\cal A}^{+}$ and ${\cal A}^{-}$, the analytic properties of the scattering amplitude are naturally satisfied. Therefore, in the high-energy limit $s \gg m^{2}$, the real and imaginary parts of the even and odd eikonals are connected through dispersion relations, expressed as \cite{qcdmodel001,qcdmodel001a,qcdmodel001b,luna001,luna001a}
\begin{eqnarray}
\chi^{+}_{_{I}}(s,b) = -\frac{2s}{\pi}\, {\cal P}\!\! \int_{0}^{\infty}ds'\,
\frac{\chi^{+}_{_{R}}(s',b)}{s^{\prime 2}-s^{2}} ,
\label{idr001}
\end{eqnarray}
and
\begin{eqnarray}
\chi^{-}_{_{I}}(s,b) = -\frac{2s^{2}}{\pi}\, {\cal P}\!\! \int_{0}^{\infty}ds'\,
\frac{\chi^{-}_{_{R}}(s',b)}{s'(s^{\prime 2}-s^{2})} ,
\label{idr002}
\end{eqnarray}
where ${\cal P}$ denotes the Cauchy principal value.

\section{the model}

In our model, similar to other minijet-type models, we assume that the eikonal function is additive with respect to the soft and semihard (SH) components of the scattering amplitude, allowing us to express it as \cite{qcdmodel001,luna001,luna001a,luna002,luna002a,luna002b,luna002c,luna003,luna003a,luna003b}
\begin{eqnarray}
\chi^{\pm}(s,b) = \chi_{\text{soft}}^{\pm}(s,b) + \chi^{\pm}_{_{\text{SH}}}(s,b).
\end{eqnarray}
             
In the QCD parton model, the crossing-odd semihard eikonal, $\chi^{-}_{_{\text{SH}}}(s,b)$, decreases rapidly with increasing $s$.
Consequently, the crossing-odd eikonal $\chi^{-}(s,b)$ receives no significant contribution from semihard processes at high energies. For this reason, it is sufficient to set $\chi_{_{\text{SH}}}(s,b)=\chi^{+}_{_{\text{SH}}}(s,b)$, leading to $\chi^{-}(s,b) = \chi^{-}_{\text{soft}}(s,b)$. 

The even part of the semihard eikonal contribution is factorized in the form\footnote{This assumption relies on a semiclassical probabilistic argument that connects the eikonal function $\chi(s,b)$ to the parton-parton cross-section $\sigma_{_{\text{QCD}}}(s)$ derived from the QCD parton model.} \cite{qcdmodel002,qcdmodel002a,qcdmodel002b,luna001,luna001a,luna002,luna002a,luna002b,luna002c,luna003,luna003a,luna003b},
\begin{eqnarray}
\chi^{+}_{_{\text{SH}}}(s,b) = \frac{1}{2}\, \sigma_{_{\text{QCD}}}(s)\, W_{\!_{\text{SH}}}(b;\nu_{_{\text{SH}}}),
\end{eqnarray}
where $\sigma_{_{\text{QCD}}}(s)$ is the usual QCD cross section for jet production, and $W_{\!_{\text{SH}}}(b;\nu_{_{\text{SH}}})$ is an overlap density for the partons at $b$ and $s$,
\begin{eqnarray}
W_{\!_{\text{SH}}}(b;\nu_{_{\text{SH}}}) = \frac{1}{2\pi}\int_{0}^{\infty}dk_{\perp}\, k_{\perp}\, J_{0}(k_{\perp}b)\,G_{A}(k_{\perp})\,G_{B}(k_{\perp}), \nonumber \\
\label{overlap001}
\end{eqnarray}
where $\nu_{_{\text{SH}}}$ is a free adjustable parameter, while $G_{A}(k_{\perp})$ and $G_{B}(k_{\perp})$ are form factors of the colliding hadrons $A$ and $B$ that depend on $\nu_{_{\text{SH}}}$ \cite{luna001,iser01,luna002a,lipari001}. We assume these form factors behave similarly to the charge dipole approximation used in the proton form factors, specifically
\begin{eqnarray}
G_{A}(k_{\perp}) = G_{B}(k_{\perp}) = \left( \frac{\nu_{_{\text{SH}}}^{2}}{k_{\perp}^{2}+\nu_{_{\text{SH}}}^{2}} \right)^{2}.
\label{form09}
\end{eqnarray}
Using this dipole form factor, we can express the overlap density as
\begin{eqnarray}
W_{\!_{\text{SH}}}(s,b) = \frac{\nu_{_{\text{SH}}}^{2}}{96\pi} (\nu_{_{\text{SH}}} b)^{3} K_{3}(\nu_{_{\text{SH}}} b),
\label{eq20}
\end{eqnarray}
where $K_{3}(x)$ is the modified Bessel function of the second kind and $\nu_{_{\text{SH}}}$ is a parameter to be determined from fits to the data. The overlap densities are normalized to satisfy $\int d^{2}b\, W(s,b) = 1$.

The leading order (LO) QCD contribution to $\sigma_{\text{tot}}$ for the inclusive process $A+B \to jets$ with $p_{T} > p_{T_{\text{min}}}$ is \cite{qcdmodel001,qcdmodel002,iser01,lipari002,forshaw01,forshaw02}
\begin{widetext}
\begin{eqnarray}
\sigma_{_{\text{QCD}}}(s) &=& \int_{p_{T_{\text{min}}}^{2}}^{s/4} dp_{T}^{2} \int_{4p_{T}^{2}/s}^{1} dx_{1}
\int_{4p_{T}^{2}/x_{1}s}^{1} dx_{2} \left[ f_{i/A}(x_{1}, Q^{2})f_{j/B}(x_{2}, Q^{2}) +
  f_{j/A}(x_{1}, Q^{2})f_{i/B}(x_{2}, Q^{2})
 \right] \nonumber \\
&\times & \left[ \frac{d\hat{\sigma}_{ij\to kl}}{dp_{T}^{2}}(\hat{t}, \hat{u}) +
  \frac{d\hat{\sigma}_{ij\to kl}}{dp_{T}^{2}}(\hat{u}, \hat{t}) \right]
  \left( 1- \frac{\delta_{ij}}{2} \right) \left( 1- \frac{\delta_{kl}}{2} \right),
\label{eq08}
\end{eqnarray}
\end{widetext}
where $\hat{s}$, $\hat{t}$, and $\hat{u}$ are the Mandelstam invariants for the parton-parton collision, with $\hat{s}+\hat{t}+\hat{u}=0$, $\hat{s}=x_{1}x_{2}s$, and $\hat{t} = -\frac{\hat{s}}{2} \left( 1 - \sqrt{1 - \frac{4p_{T}^{2}}{\hat{s}}}  \right)$. Here $p_{T_{\text{min}}}$ is the minimal momentum transfer in the semihard scattering, $x_{1}$ and $x_{2}$ are the fractions of the momenta of the parent hadrons $A$ and $B$ carried by the partons $i$, $j$, $k$, and $l$ with $i,j,k,l = q, \bar{q}, g$, $\frac{d\hat{\sigma}_{ij\to kl}}{dp_{T}^{2}}$ is the differential cross section for $ij\to kl$ scattering, and $f_{i/A}(x_{1},|\hat{t}|)$ $\left[ f_{j/B}(x_{2},|\hat{t}| \right]$ is the parton $i$ $\left[ j \right]$ distribution in the hadron $A$ $\left[ B \right]$. We adopt $Q^{2} = p_{T}^{2}$.

As $x\to 0$, the gluon distribution becomes dominant, making it essential for the parton-parton scattering processes used in the computation of $\chi_{_{\text{SH}}}(s,b)$ to involve at least one gluon in the initial state. Therefore, the calculation of $\sigma_{_{\text{QCD}}}(s)$ focuses on the processes $gg \to gg$, $qg\to qg$, $\bar{q}g\to \bar{q}g$, and  $gg\to \bar{q}q$. For instance, at  $\sqrt{s} = 7$ TeV and $p_{T_{\text{min}}} = 1.3$ GeV, these processes collectively account for at least 99.1\% of $\sigma_{_{\text{QCD}}}(s)$ when utilizing fine-tuned PDFs.

To ensure consistency, the LO expression in Eq. (\ref{eq08}) must be evaluated using LO parton–parton cross sections together with LO PDFs. Since LO jet production is highly sensitive to the factorization and renormalization scales, it is essential to test the results against different scale choices. However, certain next-to-leading-order (NLO) corrections can be incorporated phenomenologically, with their effects (particularly on the normalization and the shape of $\sigma_{_{\text{QCD}}}(s)$) effectively reproduced through variations of parameters such as the cutoff $p_{T_{\text{min}}}$, the overall normalization factor, and the parton distribution functions. While not equivalent to a full NLO treatment, this strategy captures part of the NLO dynamics while preserving the analytical simplicity of the model. Therefore, in our implementation of (\ref{eq08}), we compute the cross section $\sigma_{_{\text{QCD}}}(s)$ using NLO parton PDFs and NLO parton-parton cross sections. This procedure allows us to preserve the simplicity of equation (\ref{eq08}) while partially reducing the strong scale dependence typically associated with purely LO calculations. As a result the scale sensitivity of $\sigma_{_{\text{QCD}}}(s)$ is mitigated, resulting in a more stable eikonal function $\chi_{_{\text{SH}}}(s,b)$ and, in turn, a more reliable prediction for the total cross section $\sigma_{\text{tot}}(s)$ across a broad energy range.

Our calculations adopt the scale choice $Q^{2} = p_{T}^{2}$, which suppresses large logarithms of the form $\ln(Q^{2}/p_{T}^{2})$ that would otherwise appear in higher-order corrections. Regarding the NLO parton-parton cross sections, since analytical expressions for these parton-level processes at NLO are unavailable, we incorporate NLO corrections through  a common $K$-factor, defined as the ratio between the NLO and the LO cross section for a given process:
\begin{eqnarray}
\frac{d\hat{\sigma}_{ij\to kl}^{NLO}}{dp_{T}^{2}} = K \frac{d\hat{\sigma}_{ij\to kl}^{LO}}{dp_{T}^{2}}.
\end{eqnarray}

The $K$-factor is, in general, scale dependent and can vary significantly with the region of phase space considered or the kinematical cuts applied. Nevertheless, despite these caveats, part of the information from an NLO calculation can still be effectively captured by a constant $K$-factor, and we adopt this simplification throughout our analyses.

Note that the choice of $p_{T_{\text{min}}}$ is somewhat arbitrary. Its value is typically the same order as $Q_{0}$, the initial scale for the DGLAP evolution of the parametric forms used in PDF determination. Nevertheless, consistent results for $\sigma_{\text{tot}}(s)$ and $\rho(s)$ are generally obtained for different values of $p_{T_{\text{min}}}$ within a certain range around $Q_{0}$. In our model, the uncertainties associated with $p_{T_{\text{min}}}$ and with the $K$-factor are absorbed into a single phenomenological parameter, ${\cal N}$, which is determined through data fitting. Consequently, the expression in equation (\ref{eq08}) is effectively multiplied by ${\cal N}$, and the even part of the semihard eikonal takes the form
\begin{eqnarray}
\chi^{+}_{_{\text{SH}}}(s,b) = \frac{1}{2}\,{\cal N}\, \sigma_{_{\text{QCD}}}(s)\, W_{\!_{\text{SH}}}(s,b).
\end{eqnarray}

The soft eikonal is required primarily to describe the lower-energy forward scattering data, as the asymptotic behavior of the hadron-hadron total cross section is dominated by parton-parton semihard collisions. For the even component of the soft eikonal, we adopt the form
\begin{eqnarray}
\chi^{+}_{\text{soft}}(s,b) = \frac{1}{2}\, W^{+}_{\!\!\text{soft}}(b;\mu^{+}_{\text{soft}})\, \left[ A + iB +\frac{C}{(s/s_{0})^{\gamma}}\, e^{i\pi\gamma/2} \right] , \nonumber \\ 
\label{soft01}
\end{eqnarray}
where $\sqrt{s_{0}}\equiv 1$ GeV, $\gamma \equiv 0.77$, and $A$, $B$, $C$, and $\mu^{+}_{\text{soft}}$ are fitting parameters. The phase factor $e^{i\pi\gamma/2}$ ensures the correct analyticity properties of the amplitude and arises from the integral dispersion relation (\ref{idr001}).

The odd soft eikonal, denoted as $\chi^{-}_{\text{soft}}(s,b)$, accounts for the differences between $pp$ and $\bar{p}p$ channels and vanishes at high energies:
\begin{eqnarray}
\chi^{-}_{\text{soft}}(s,b) &=& \frac{1}{2}\, W^{-}_{\!\!\text{soft}}(b;\mu^{-}_{\text{soft}})\,D\, \frac{e^{-i\pi/4}}{\sqrt{s/s_{0}}},
\label{softminus}
\end{eqnarray}
where $\mu^{-}_{\text{soft}}$ is fixed at 0.5 GeV. Here $D$ represents the strength of the odd term and constitutes another fitting parameter. The phase factor $e^{-i\pi/4}$ ensures the proper analyticity properties of the expression, as derived from the dispersion relation (\ref{idr002}).

All soft parameters, both free and fixed, show minimal statistical correlation with the semihard parameters. In addition, the fixed parameters $\sqrt{s_{0}}$, $\gamma$, and $\mu^{-}_{\text{soft}}$ are consistent with the values reported in previous studies \cite{iser01}.

The soft form factors are assumed to have the same structure as the semihard form factor, namely
\begin{eqnarray}
W_{\text{soft}}(b;\mu^{i}_{\text{soft}}) = \frac{(\mu^{i}_{\text{soft}})^{2}}{96\pi} (\mu^{i}_{\text{soft}} b)^{3} K_{3}(\mu^{i}_{\text{soft}} b) ,
\label{chDGM.19}
\end{eqnarray}
with $i=+,-$, where $\mu^{+}_{\text{soft}}$ and $\mu^{-}_{\text{soft}}$ correspond to the fitting parameters already defined in Eqs. (\ref{soft01}) and (\ref{softminus}).

Finally, the forward quantities $\sigma_{\text{tot}}(s)$ and $\rho(s)$ can be expressed in terms of the eikonal function $\chi (s,b)$. From the optical theorem,
\begin{eqnarray}
  \sigma_{\text{tot}}(s) = 4\pi\, \textnormal{Im}\,{\cal A}(s,t=0),
\end{eqnarray}
and the total cross section reads
\begin{eqnarray}
  \sigma_{\text{tot}}(s) = 4\pi   \int_{_{0}}^{^{\infty}}   \!\!  b\,   db\,
[1-e^{-\chi_{_{R}}(s,b)}\cos \chi_{_{I}}(s,b)] ,
\end{eqnarray}
while the $\rho$ parameter, which represents the ratio of the real to imaginary parts of the forward scattering amplitude,
\begin{eqnarray}
  \rho(s) = \frac{\textnormal{Re}\,{\cal A}(s,t=0)}{\textnormal{Im}\,{\cal A}(s,t=0)},
\end{eqnarray}
is given by
\begin{eqnarray}
  \rho(s) = \frac{-\int_{_{0}}^{^{\infty}}   \!\!  b\,  
db\, e^{-\chi_{_{R}}(s,b)}\sin \chi_{_{I}}(s,b)}{\int_{_{0}}^{^{\infty}}   \!\!  b\,  
db\,[1-e^{-\chi_{_{R}}(s,b)}\cos \chi_{_{I}}(s,b)]}.
\end{eqnarray}

\section{Results and Conclusions}

We perform global fits to forward ($t = 0$) scattering data in the energy range from $\sqrt{s}_{\text{min}} = 10$ GeV up to LHC energies. The datasets include those compiled and analyzed by the Particle Data Group (PDG) \cite{PDG} (10 GeV $\leq \sqrt{s} \leq$ 1.8 TeV), as well as the measurements at LHC from the ATLAS/ALFA \cite{atlas001,atlas001a,atlas01} and TOTEM \cite{TOTEM001,TOTEM002,Tot-2,TOTEM004,Tot-1,TOTEM007,TOTEM010,antchev01} Collaborations. Specifically, we fit the total cross sections, $\sigma_{\text{tot}}^{pp}$ and $\sigma_{\text{tot}}^{\bar{p}p}$, along with the ratios of the real to imaginary part of the forward scattering amplitude, $\rho^{pp}$ and $\rho^{\bar{p}p}$. The statistical and systematic uncertainties are combined in quadrature.

Within the portion of the dataset measured at the LHC, the measurements of $\sigma_{\text{tot}}^{pp}$ reveal a notable tension between the ATLAS/ALFA and TOTEM results at $\sqrt{s} =$ 7, 8, and 13 TeV \cite{tension1,tension2,tension3,tension4,tension5,lkmr01,lkmr01a,lkmr01b,lkmr01c}. For example, the TOTEM measurement of $\sigma_{\text{tot}}^{pp}$ at $\sqrt{s} = 7$ TeV, $\sigma_{\text{tot}}^{pp} = 98.58 \pm 2.23$ mb \cite{TOTEM002}, differs from the ATLAS/ALFA value at the same energy, $\sigma_{\text{tot}}^{pp} = 95.35 \pm 1.36$ mb \cite{atlas001}, by 1.4 $\sigma$ (assuming uncorrelated uncertainties \cite{broilo2021a}). Similarly, at $\sqrt{s} = 8$ TeV, the ATLAS/ALFA total cross section, $\sigma_{\text{tot}}^{pp} = 96.07 \pm 0.92$ mb \cite{atlas001a}, differs from the lowest TOTEM measurement, $\sigma_{\text{tot}}^{pp} = 101.5 \pm 2.1$ mb \cite{TOTEM007}, by 2.6 $\sigma$. A comparable level of disagreement is also present at 13 TeV. This persistent tension  implies different possible scenarios for the rise of the total cross section.
Regarding the measurements of the $\rho^{pp}$ parameter, TOTEM has reported values at 7, 8, and 13 TeV, while ATLAS/ALFA has only a single measurement at 13 TeV. Unlike the discrepancies observed in the $\sigma_{\text{tot}}^{pp}$ data, the $\rho^{pp}$ measurements at 13 TeV from ATLAS/ALFA ($\rho^{pp} = 0.098 \pm 0.011$ \cite{atlas01}) and TOTEM ($\rho^{pp} = 0.09 \pm 0.01$ and $\rho^{pp} = 0.10 \pm 0.01$ \cite{antchev01}) are fully consistent with each other.

To effectively investigate the tension between the TOTEM and ATLAS/ALFA results, we perform global fits to the $pp$ and $\bar{p}p$ forward data using two distinct datasets: one that includes only the TOTEM measurements, and another that includes only the ATLAS/ALFA results. 

The two resulting data ensembles are defined as follows: \\

{\bf Ensemble A}: $\left. \sigma_{\text{tot}}^{pp,\bar{p}p}\right|_{\footnotesize \textnormal{PDG}}$ + $\left. \rho^{pp,\bar{p}p}\right|_{\footnotesize \textnormal{PDG}}$ + $\left. \sigma_{\text{tot}}^{pp}\right|_{\footnotesize \textnormal{ATLAS/ALFA}}$ + $\left. \rho^{pp}\right|_{\footnotesize \textnormal{ATLAS/ALFA}}$ \\

{\bf Ensemble T}: $\left. \sigma_{\text{tot}}^{pp,\bar{p}p}\right|_{\footnotesize \textnormal{PDG}}$ + $\left. \rho^{pp,\bar{p}p}\right|_{\footnotesize \textnormal{PDG}}$ + $\left. \sigma_{\text{tot}}^{pp}\right|_{\footnotesize \textnormal{TOTEM}}$ + $\left. \rho^{pp}\right|_{\footnotesize \textnormal{TOTEM}}$ \\

In the {\bf Ensemble A}, the $\left. \sigma_{\text{tot}}^{pp}\right|_{\footnotesize \textnormal{ATLAS/ALFA}}$ dataset encloses measurements of the total cross section at $\sqrt{s} =$ 7 \cite{atlas001}, 8 \cite{atlas001a}, and 13 \cite{atlas01} TeV. The $\left. \rho^{pp}\right|_{\footnotesize \textnormal{ATLAS/ALFA}}$ dataset consists of the $\rho$ measurement at $\sqrt{s}=13$ TeV \cite{atlas01}.

In the {\bf Ensemble T}, the $\left. \sigma_{\text{tot}}^{pp}\right|_{\footnotesize \textnormal{TOTEM}}$ dataset includes measurements of the total cross section at $\sqrt{s} =$ 2.76 \cite{totem276}, 7 \cite{TOTEM001,TOTEM002,Tot-2}, 8 \cite{TOTEM004,Tot-1,TOTEM007}, and 13 \cite{TOTEM010,antchev01} TeV. The $\left. \rho^{pp}\right|_{\footnotesize \textnormal{TOTEM}}$ dataset comprises $\rho$ measurements at $\sqrt{s}=$ 7 \cite{Tot-2}, 8 \cite{Tot-1}, and 13 \cite{antchev01} TeV.

In this work, we employ two sets of parton distribution functions at NLO: NNPDF4.0 \cite{NNPDF} and CT18 \cite{CT18}, all available through LHAPDF6 \cite{LHAPDF}. The NNPDF family is constructed using a machine-learning framework based on neural networks, which allows for highly flexible parametrizations of the PDFs at the initial scale $Q_0$ with minimal theoretical bias. These sets incorporate a wide range of collider data from HERA, the Tevatron, and the LHC, with the more recent versions introducing an improved architecture, more sophisticated uncertainty estimation, and an expanded dataset that includes vector boson and jet production at 7, 8, and 13 TeV. These additions result in significantly improved constraints on the gluon distribution, particularly in the small-$x$ region relevant for high-energy scattering. The CT18 PDFs represent the latest global analysis from the CTEQ-TEA collaboration, featuring an updated parametrization at the initial scale $Q_0$ based on Bernstein polynomials, which provides enhanced flexibility in the DGLAP evolution. Compared to its predecessor CT14 \cite{CT14}, the CT18 set incorporates more precise LHC measurements of inclusive vector boson and jet production, leading to a substantial reduction in the gluon PDF uncertainties, especially at intermediate and small values of $x$. We also carried out global fits using a third NLO PDF set, MSHT20 \cite{MSHT20}, which incorporates important methodological advances. These include an improved treatment of heavy-quark masses, a more refined handling of correlated experimental uncertainties, and an extensive inclusion of recent LHC data. As a result, the gluon distribution is better constrained across the full kinematic range, particularly at small $x$, which is essential for gluon-dominated processes such as minijet production. However, the growth rate of $\sigma_{_{\text{QCD}}}(s)$ obtained with MSHT20 is excessively steep, leading to statistically poor fits for both $\sigma_{\text{tot}}$ and $\rho$.

The fitted parameter values are listed in Table I (Ensemble A) and Table II (Ensemble T). The fits were obtained through a $\chi^{2}$ minimization procedure, with confidence regions defined by the interval $\chi^{2}-\chi^{2}_{\text{min}}$ corresponding to the 90\% confidence level. In our case, this corresponds to $\chi^{2}-\chi^{2}_{\text{min}} = 12.02$ for seven free parameters. The minimum value, $\chi^{2}_{\text{min}}$, follows a $\chi^{2}$ distribution with $\nu$ degrees of freedom. As a convergence requirement, we retained only fits that yielded positive-definite covariance matrices. The resulting $\chi^{2}/\nu$ values were computed for 158 degrees of freedom in the case of Ensemble A, and 168 in the case of Ensemble T.

It is worth emphasizing that only two free parameters are associated with the semihard term, which drives the rise of the total cross section and determines the high-energy behavior. The remaining parameters describe the low-energy region, where differences between the $pp$ and $\bar{p}p$ channels are relevant. As shown in Tables~I and II, the number of soft parameters can be further reduced: with NNPDF4.0, the parameter $C$ is compatible with zero for both ensembles, and the same occurs for Ensemble~A with CT18. Setting $C=0$ in these cases yields fits of comparable quality, with only minor shifts in the other soft parameters.
Given that the high-energy behavior of $\sigma_{\text{tot}}^{pp,\bar{p}p}$ and $\rho^{pp,\bar{p}p}$ depends solely on the semihard parameters $\mathcal{N}$ and $\nu_{_{\text{SH}}}$, no additional attempts to reduce the soft sector were undertaken.

\begin{table}
\centering
\caption{Fitted parameter values obtained from the global analysis for Ensemble A. The best-fit results correspond to $p_{T_{\text{min}}} = 1.1$ GeV for CT18 and $p_{T_{\text{min}}} = 1.3$ GeV  NNPDF4.0.}
\begin{ruledtabular}
\begin{tabular}{ccc}
  & CT18 &  NNPDF4.0 \\
 \hline
  & $p_{Tmin}=1.1\,\mathrm{GeV}$ & $p_{Tmin}=1.3\,\mathrm{GeV}$ \\
 \hline \\[-2.2ex]
 $\mathcal{N}$ & $1.83\pm 0.46$ &  $1.53\pm 0.57$ \\
 $\nu_{_{\text{SH}}}$ [GeV] & $1.418\pm 0.084$ &  $1.12\pm 0.12$ \\
 $A$ $[\mathrm{GeV}^{-2}]$ & $(2.1\pm 2.0)\times 10^3$ & $(3.9\pm 4.9)\times 10^{2}$ \\
 $B$ $[\mathrm{GeV}^{-2}]$ & $-76\pm 148$ &  $-14\pm 26$ \\
 $C$ $[\mathrm{GeV}^{-2}]$ & $(31\pm 36)\times10^{3}$ &   $(4.5\pm 7.6)\times 10^{3}$ \\
 $\mu_{soft}^{+}$ [GeV] & $2.55\pm 0.39$ &   $1.78\pm 0.67$ \\
 $D$ $[\mathrm{GeV}^{-2}]$ & $150\pm 11$ &   $147\pm 12$ \\
 \hline \\[-2.2ex]
 $\nu$ & $158$  & $158$ \\
 $\chi^{2}/\nu$ & $1.10$  & $1.07$ \\
\end{tabular}
\end{ruledtabular}
\end{table}

\begin{table}
\centering
\caption{Fitted parameter values obtained from the global analysis for Ensemble T. The optimal values of $p_{T_{\text{min}}}$ are the same as those in Table I, namely $p_{T_{\text{min}}} = 1.1$ GeV for CT18 and $p_{T_{\text{min}}} = 1.3$ GeV  NNPDF4.0.}
\begin{ruledtabular}
\begin{tabular}{ccc}
  & CT18 &  NNPDF4.0 \\
 \hline
  & $p_{Tmin}=1.1\,\mathrm{GeV}$ & $p_{Tmin}=1.3\,\mathrm{GeV}$ \\
 \hline \\[-2.2ex]
 $\mathcal{N}$ & $1.49\pm 0.50$ & $1.42\pm 0.49$ \\
 $\nu_{_{\text{SH}}}$ [GeV] & $1.32\pm 0.12$ & $1.13\pm 0.11$ \\
 $A$ $[\mathrm{GeV}^{-2}]$ & $(2.38\pm 0.24)\times 10^3$ & $(6.3\pm 12)\times 10^{2}$\\
 $B$ $[\mathrm{GeV}^{-2}]$ & $-79\pm 131$ & $-23\pm 60$ \\
 $C$ $[\mathrm{GeV}^{-2}]$ & $(34\pm 11)\times10^{3}$ & $(8.2\pm 20)\times 10^{3}$ \\
 $\mu_{soft}^{+}$ [GeV] & $2.584\pm 0.049$ & $2.02\pm 0.91$ \\
 $D$ $[\mathrm{GeV}^{-2}]$ & $149\pm 11$ & $148\pm 12$ \\
 \hline \\[-2.2ex]
 $\nu$ & $168$  & $168$ \\ 
 $\chi^{2}/\nu$ & $1.18$  & $1.11$ \\
\end{tabular}
\end{ruledtabular}
\end{table}   

The results of the fits to $\sigma_{\text{tot}}$ and $\rho$ for the $pp$ and $\bar{p}p$ channels, together with the experimental data, are shown in Figs.~1 and 2. In Fig. 1 (Ensemble A), the ATLAS/ALFA measurements of $\sigma_{\text{tot}}^{pp}$ are well described. For $\rho^{pp}$, however, the result obtained with CT18 fails to reproduce the 13 TeV data, whereas the result with NNPDF4.0 intersects the experimental uncertainty band. Specifically, at $\sqrt{s}=13$ TeV the CT18 fit yields $\sigma_{\text{tot}}=103.2$ mb and $\rho^{pp}=0.113$, while the NNPDF4.0 fit gives $\sigma_{\text{tot}}=102.7$ mb and $\rho^{pp}=0.105$.
In Fig. 2 (Ensemble T), the TOTEM measurements of $\sigma_{\text{tot}}$ are also well described. For $\rho$, the CT18 result again fails to describe the 13 TeV data, while the NNPDF4.0 result lies close to the upper edge of the experimental uncertainty band. At $\sqrt{s}=13$ TeV the CT18 fit yields $\sigma_{\text{tot}}^{pp}=107.8$ mb and $\rho^{pp}=0.120$, whereas the NNPDF4.0 fit gives $\sigma_{\text{tot}}^{pp}=107.1$ mb and $\rho^{pp}=0.111$.

For comparison, our results can be directly confronted with the ATLAS/ALFA measurement of $\sigma_{\text{tot}}^{pp}$ at $\sqrt{s}=13$ TeV, $\sigma_{\text{tot}}^{pp}=104.68\pm1.08$, the average of the three TOTEM values, $\overline{\sigma}_{\text{tot}}=110.1\pm2.0$, and the average of the three measurements of $\rho^{pp}$ from ATLAS/ALFA and TOTEM, $\overline{\rho}=0.096\pm0.006$. Although the global fits obtained with CT18 and NNPDF4.0 are statistically satisfactory, and the predicted values of $\rho^{pp}$ at 13 TeV are of the same order of magnitude as the averaged value $\overline{\rho}$, a noticeable discrepancy remains between our results at 13 TeV and the central values reported by the ATLAS/ALFA and TOTEM Collaborations.
As discussed earlier in this work, such a deviation may indicate the need to include an odd-under-crossing contribution in the semihard sector. Within this framework, both even and odd components of the scattering amplitude become relevant at high energies, providing phenomenological support for the existence of a color-singlet state composed of three reggeized gluons, the QCD Odderon.

The development of an odd component in the scattering amplitude that survives at asymptotically high energies, along with a systematic study of the sensitivity of the results to the unitarization scheme \cite{tension2,tension3,ryskin_luna01}, is currently in progress.


\section*{Acknowledgments}

The authors thank A.V. Giannini and M.G. Ryskin for carefully reading the manuscript and for valuable discussions.
This research was partially supported by the Conselho Nacional de Desenvolvimento Cient\'{\i}fico e Tecnol\'ogico (CNPq).

\begin{figure*}
\label{figpaperluna01}
\begin{center}
\includegraphics[height=.9\textheight]{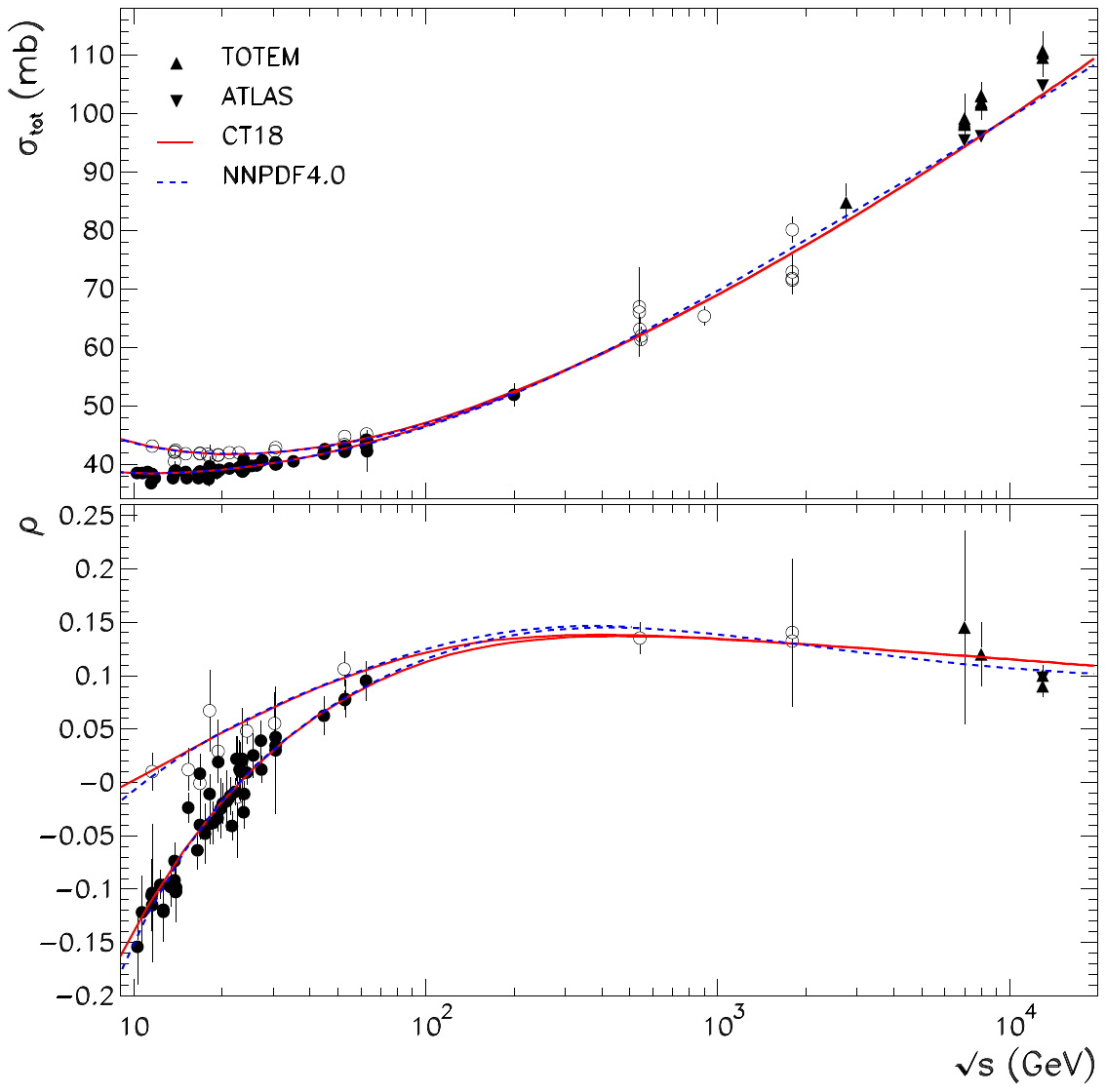}
\vspace{-1.6cm}  
\caption{Description of $\sigma_{\text{tot}}(s)$ and $\rho(s)$ obtained from the global fit to Ensemble A. The solid and and dashed curves correspond to the results using CT18 and NNPDF4.0, respectively.}
\end{center}
\end{figure*}

\begin{figure*}
\label{figpaperluna01}
\begin{center}
\includegraphics[height=.9\textheight]{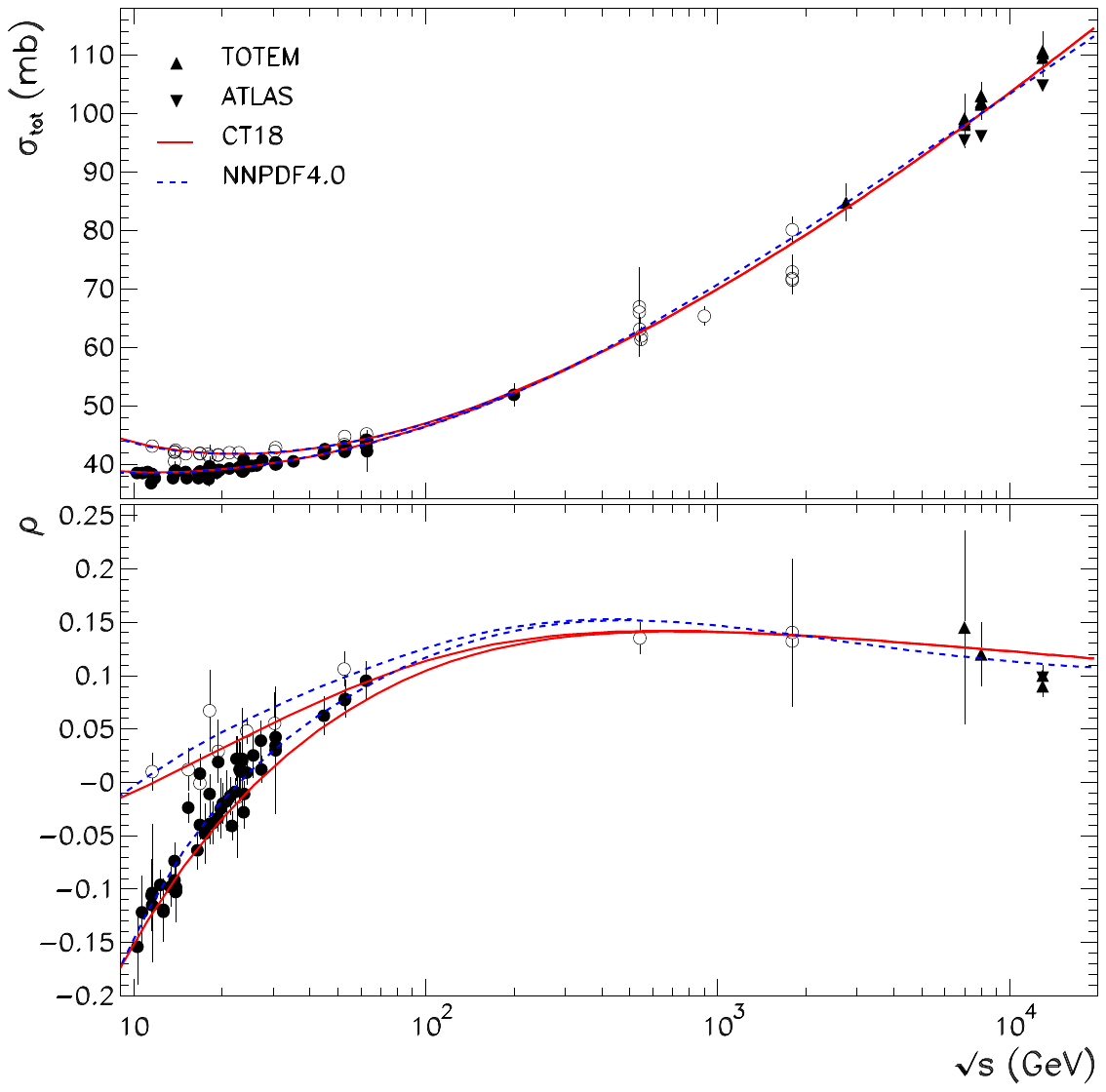}
\vspace{-1.6cm}  
\caption{Description of $\sigma_{\text{tot}}(s)$ and $\rho(s)$ obtained from the global fit to Ensemble T. The solid and and dashed curves correspond to the results using CT18 and NNPDF4.0, respectively.}
\end{center}
\end{figure*}

\end{document}